\documentclass[journal]{IEEEtran}
\usepackage{graphicx}
\usepackage{cite}
\usepackage{latexsym}
\usepackage{pifont}
\usepackage{textcomp}
\usepackage{marvosym}
\usepackage{amssymb}
\usepackage{subfigure}
\usepackage{amsmath, amsthm, amssymb}
\usepackage{algpseudocode}
\usepackage{algorithm}
 \usepackage{multibib}
 \usepackage[english]{babel}

\begin{document}

\sloppy

\title{Distributed Voting/Ranking with Optimal Number of States per Node}

\author{
  \IEEEauthorblockN{Saber Salehkaleybar{\em, Member, IEEE}, Arsalan Sharif-Nassab, and S. Jamaloddin Golestani{\em, Fellow, IEEE}}
  \\
  \IEEEauthorblockA{Dept. of Electrical Engineering, Sharif University of Technology, Tehran, Iran\\
    Emails: saber\_saleh@ee.sharif.edu, sharifnassab@ee.sharif.edu, golestani@ieee.org} 
}

\maketitle
\begin{abstract}
Considering a network with $n$ nodes, where each node initially votes for one (or more) choices out of $K$ possible choices, we present a Distributed Multi-choice Voting/Ranking (DMVR) algorithm to determine either the choice with maximum vote (the voting problem) or to rank all the choices in terms of their acquired votes (the ranking problem). The algorithm consolidates node votes across the network by updating the states of interacting nodes using two key operations; the union and the intersection. The proposed algorithm is simple, independent from network size, and easily scalable in terms of the number of choices $K$, using only $K\times 2^{K-1}$ nodal states for voting, and $K\times K!$ nodal states for ranking. We prove the number of states to be optimal in the ranking case; this optimality is conjectured to also apply to the voting case. The time complexity of the algorithm is analyzed in complete graphs. We show that the time complexity for both ranking and voting is $O(\log(n))$ for given vote percentages, and is inversely proportional to the minimum of the vote percentage differences among various choices.   
\end{abstract}

\theoremstyle{definition}
\newtheorem{mydef}{Definition}
\newtheorem{myex}{Example}
\newtheorem{myrm}{Remark}
\theoremstyle{theorem}
\newtheorem{mylm}{Lemma}
\newtheorem{myth}{Theorem}
\section{Introduction}
One of the key building blocks in distributed function computation is \textquotedblleft Majority Voting\textquotedblright. It can be employed as a subroutine in many network applications such as target detection in sensor networks \cite{katenka2008local,ridout2013improved}, distributed hypothesis testing \cite{rhim2014distributed}, quantized consensus \cite{kar2010distributed}, voting in distributed systems \cite{angluin2006computation}, and molcular nanorobots \cite{chen2013programmable}. In the distributed majority voting, each node chooses a candidate from a set of choices and the goal is to determine the candidate with the majority vote by running a distributed algorithm. As an example in target detection \cite{katenka2008local}, wireless sensors combine their binary decisions about the presence of a target through majority voting, and send a report to the fusion center if the majority is in favor of presence.

The majority voting problem for the binary case has been extensively studied in cellular automata (CA) literature. In \cite{land1995no}, it has been shown that there is no synchronous deterministic two-state automaton that can solve binary voting problem in a connected network. Several two-state automata have been proposed for the ring topology \cite{peleg1997local,mustafa2001majority}, the most successful of which can get the correct result in nearly $83\%$ of initial configurations of selected votes \cite{de1992gacs}. In order to circumvent the impossibility result  of \cite{land1995no}, asynchronous and probabilistic automata have also been presented in CA community \cite{hassin2001distributed,fuks2002nondeterministic,tomassini2002evolution}. However, none of them can obtain the correct result with probability one \cite{Benezit2011}. Using a different approach, binary voting problem can be solved by a randomized gossip algorithm \cite{Boyd2006} that computes the average of initial node values. The drawback of this approach is that the number of required states in its quantized version \cite{Kashyap2006} grows linearly in terms of the network size \cite{Benezit2011}.

In applying gossip algorithms to the implementation of binary majority voting, node does not need to come up with the exact average of the node values; it suffices to determine interval to which the average node values belongs. From this observation, B\'en\' ezit et al. \cite{benezit2009interval} proposed an elegant solution based on an automaton, with the state space $\{0,0.5^-,0.5^+,1\}$, which resembles the idea in \cite{Kashyap2006}. The initial state of nodes voting for $\mbox{\textquotedblleft 0\textquotedblright\space or \textquotedblleft 1\textquotedblright}$ is $0$ or $1$, respectively. When two neighbor nodes get in contact with each other, they exchange their states and update them according to a transition rule. It can be shown that the states of all nodes would be in the set $\{0,0.5^-\}$ at the end of the algorithm, if the choice \textquotedblleft 0\textquotedblright\space is in majority. Otherwise, the state of all nodes would belong to the set $\{0.5^+,1\}$. In \cite{Benezit2011}, a Pairwise Asynchronous Graph Automata (PAGA) has been used to extend the above idea to the multiple choice voting problem, and sufficient conditions for convergence are stated. This approach results in a 15-state automaton and a 100-state automaton for the ternary and quaternary voting problems, respectively. For majority voting with more than four choices, pairwise and parallel comparison among the choices, has been proposed \cite{Benezit2011}, requiring $\Theta(2^{K(K-1)})$ number of states in terms of the number of choices, $K$. At the end, authors posed a few open problems. One of the main problems is whether voting automata exist for any number of multiple choices without running multiple binary or ternary voting automata in parallel? Furthermore, what is the minimum number of states of a possible solution?

In more recent works \cite{becchetti2014simple,becchettiplurality,jung2012distributed,babaee2013distributed}, it has been shown that the majority vote can be obtained with high probability if the initial votes are sufficiently biased to the majority or the network size is large enough. However, none of these works can guarantee convergence to the correct result.

A generalization of the distributed voting problem, is the distributed ranking problem in which the goal is to rank all the $K$ choices in terms of the number of votes, each get from different network nodes \cite{jung2012distributed}. In this paper, we propose a Distributed Multi-choice Voting/Ranking (DMVR) Algorithm for solving the majority voting and ranking problems in general networks. The proposed algorithm may also be applied where each node is allowed to vote for more than one choice. Our main contributions are summarized as follows:
\begin{itemize}
\item Our proposed DMVR algorithm provides a simple and easily scalable approach for distributed voting and ranking that works for any number $K$ of choices, requiring $K\times 2^{K-1}$ and $K\times K!$ number of states for the voting and ranking problems, respectively. For instance, the number of required states is $12$ for ternary voting, and $32$ for quaternary voting, compared to respectively $15$ and $100$ states in the case of PAGA algorithm \cite{Benezit2011}. Furthermore, unlike the randomized gossip algorithms \cite{Kashyap2006,Boyd2006}, the number of states is independent from the network size. 

\item We establish a lower bound on the number of required states in any ranking algorithm, and show that the DMVR algorithm achieves this bound. Compared to the existing algorithms, the state of the DMVR algorithm can be encoded by roughly $\Theta(K\log(K))$ bits.

\item In complete graphs, we analyze the time complexity of the DMVR algorithm for the ranking problem. We will show how the time complexity is related to the percentage of nodes voting for different choices. Besides, we propose a modification for speeding up the DMVR algorithm for the majority voting problem.
\end{itemize}

The remainder of this paper is organized as follows: In Section II, the DMVR algorithm for majority voting and ranking is described. Section III studies the convergence of the DMVR algorithm. Furthermore, the number of states of the DMVR algorithm is analyzed in both cases of voting and ranking. Section IV is devoted to analyze the time complexity of the DMVR algorithm in complete graphs. In Section V, simulation results are provided. Finally, we conclude the paper in Section VI.
\section{The Distributed Multi-choice Voting/Ranking (DMVR) algorithm}
\subsection{Problem Statement}
Consider a network with $n$ nodes. The topology of the network is represented by a connected undirected graph, $G=(V,E)$, with the vertex set $V=\{1,...,n\}$, and the edge set $E\subseteq V\times V$, such that $(i,j)\in E$ if and only if nodes $i$ and $j$ can communicate directly. Furthermore, it is assumed that each node is equipped with a local clock which ticks according to a Poisson process with rate one. Initially, each node $i$ chooses a choice from a set of $K$ choices $\mathcal{C}=\{c_1,\cdots,c_K\}$. Let $\# c_k$ be the number of nodes that select the choice $c_k$ and $\rho_k\triangleq \frac{\#c_k}{n}$. In the majority voting problem, the goal is to find the choice $c_k$ in majority, i.e. the choice $c_k$ satisfying $\# c_k\geq\# c_j, j\in \{1,\cdots,K\}$. In the ranking problem, the desired output is a permutation, $[\pi_1,\cdots,\pi_K]$ of $\mathcal{C}$ such that $\#\pi_k\geq\#\pi_{k+1}, \forall k$.
\subsection{Description of the DMVR algorithm}
A value set $v_i(t)$ is associated with each node $i$ at time $t$. At $t=0$, the only member of $v_i(0)$ is the selected choice of node $i$. In the process of the algorithm, $v_i(t)$ always remains a subset of $\mathcal{C}$. The algorithm essentially performs two function. One function of the algorithm deals with consolidating node choices across the network. This function utilizes two key operations, the union and the intersection, in order to update the value sets $v_i(t)$ and $v_i(t)$ of nodes $i$ and $j$, when they interact. The second part of the algorithm has to do with disseminating the consolidated result of part one throughout the network. For reasons to be clarified later, the above two functions of the algorithm are executed in parallel, not sequentially. The second function of the algorithm operates on a collection of sets $m_{i,k}(t)$, $1\leq k\leq K$, at each node $i$. We collectively refer to the sets $m_{i,k}(t)$, $1\leq k\leq K$, of node $i$ as memory of it. Each $m_{i,k}(t)$ is a subset of $\mathcal{C}$. Unlike the dissemination function, the consolidation function is identical for the voting and ranking. In the following, we describe the dissemination function for the more general case, i.e. for the ranking case, the dissemination for the voting case being a special and simplified version of it.

When node $i$'s clock ticks at time $t$, it chooses one of the neighbor nodes, say node $j$, randomly. Then, nodes $i$ and $j$ update their value sets and memories according to the following transition rules:
\begin{align}
\nonumber &\mbox{\bf{Consolidation Function:}}
\nonumber\\&\resizebox{1.06\hsize}{!}{$  \begin{cases}
v_i(t^+):=  v_i(t) \cup v_j(t),\;  v_j(t^+):= v_i(t)\cap v_j(t), \mbox{ if  } |v_i(t)|\leq |v_j(t)|,\\
\nonumber v_i(t^+):= v_i(t) \cap v_j(t),\;  v_j(t^+):= v_i(t)\cup v_j(t), \mbox{ Otherwise}.
\end{cases}$}
\\
\nonumber &\mbox{\bf{Dissemination Function:}}
\\ & \qquad m_{i,|v_i(t^+)|}(t^+):=v_i(t^+), \quad m_{j,|v_j(t^+)|}(t^+):=v_j(t^+),
\label{eqrule}
\end{align}
where there is no memory updating if $v_i(t^+)=\emptyset$. Furthermore, we have: $m_{i,k}(t^+)=m_{i,k}(t)$ for $\forall k\neq |v_i(t^+)|$, $\forall i\in\{1,\cdots,n\}$.

\begin{algorithm}[!t]
\caption{The Distributed Multi-choice Voting/Ranking Algorithm} 
\label{Voting} 
\begin{algorithmic}[1]
\State Initialization: $v_i(0)=$ Initial vote of node $i$, $m_{i,k}(0)=\emptyset, 1\leq k\leq K$, $\forall i \in \{1,\cdots,n\}$.
\If {Node $i$'s clock ticks at time $t$}  
\State Node $i$ contacts with a random neighbor node, say node $j$: 
\If {$|v_i(t)|\leq |v_j(t)|$} 
\State $v_i(t^+):= v_i(t) \cup v_j(t),\;  v_j(t^+):= v_i(t)\cap v_j(t)$.
\Else
\State $v_i(t^+):= v_i(t) \cap v_j(t),\;  v_j(t^+):= v_i(t)\cup v_j(t)$.
\EndIf
\State $m_{i,|v_i(t^+)|}(t^+):=v_i(t^+), m_{j,|v_j(t^+)|}(t^+):=v_j(t^+).$
\EndIf

\end{algorithmic} 
\end{algorithm}

When the algorithm converges\footnote{In Section III, Theorem 1, we will describe when the algorithm eventually converges to the correct result.}, each node $i$ can obtain the correct ranking as follows\footnote{ The ``$\backslash$" is the set-theoretic difference operator, i.e. $A\backslash B= \{x: x\in A, x\not\in B\}$ for any sets $A$ and $B$.}:
\begin{equation}
\pi_k=
\begin{cases}
m_{i,k}(t)\backslash m_{i,k-1}(t), &\mbox{if } k>1,\\
m_{i,1}(t), & \quad k=1.
\end{cases}
\label{eq2}
\end{equation}

In the case of the majority voting problem, it suffices to keep the memory $m_{i,1}(t)$ at each node $i$. We denote $m_{i,1}(t)$ by $m_i(t)$ when the DMVR algorithm is executed for the majority voting problem. The description of the DMVR Algorithm is given in Algorithm 1.

Suppose that $|v_i(t)|\leq |v_j(t)|$. It is not difficult to show that the updating rule in (\ref{eqrule}) has the following properties:
\begin{itemize}
\item Define the size of choice $c_k$ as: $|\{i\mid c_k\in v_i(t)\}|$. The size of every choice $c_k$ is preserved during the updates.
\item We have $v_j(t^+)\subseteq v_i(t^+)$. If $v_i(t)\subseteq v_j(t)$, the two nodes just exchange their value sets.
\item The quantity $|v_i(t)|^2+|v_j(t)|^2$ strictly increases if $v_i(t)\not\subseteq v_j(t)$. Otherwise, it remains unchanged.
\end{itemize}

\section{Convergence Analysis}
In this section, we will show that the DMVR algorithm converges to the correct solution for the majority voting and ranking problems. First, we study how value sets consolidate and get in a convergence set, by defining a Lyapunov function. Then, we discuss how memory updating can disseminate the correct result in parallel to value set updating. Next, we merge value sets and memories of the DMVR algorithm in order to reduce memory usage in both majority voting and ranking problems. At the end, we prove that the proposed implementation is optimal in terms of the required number of states for the ranking problem.  
\subsection{Consolidation of Value Sets}
In this part, we analyze how value sets consolidate in the network until the state of the system gets in a convergence set.

\begin{mydef}
Let the network state vector at time $t$ be defined as $X(t)=[v_1(t),\cdots,v_n(t)]$. The set of all state vectors $X(t)=[v_1(t),\cdots,v_n(t)]$ with the following property is called the convergence set and is denoted by $\mathcal{X}_0$:
\begin{equation}
|v_i(t)|\leq |v_j(t)| \Longrightarrow v_i(t) \subseteq v_j(t), \forall i,j\in\{1,\cdots,n\}.
\end{equation}
 
\label{def1}
\end{mydef}

\begin{myex}
Consider a network of $n=8$ nodes and three possible choices $\mathcal{C}=\{c_1,c_2,c_3\}$. Assume that $X(0)=[\{c_1\},\{c_1\},\{c_2\},\{c_3\},\{c_1\},\{c_3\},\{c_2\},\{c_1\}]$. The state vector $X(t)=[\{c_1\},\emptyset,\{c_1,c_2,c_3\},\emptyset,\emptyset, \{c_1,c_2\},\{c_1,c_3\},\emptyset]$ cannot be in the set $\mathcal{X}_0$ since $\{c_1,c_2\}\not\subseteq \{c_1,c_3\}$. However, the state vector $X(t)=[\{c_1,c_2,c_3\},\emptyset,\{c_1,c_2,c_3\},\emptyset,\emptyset, \{c_1\},\{c_1\},\emptyset]$ is a member of the convergence set.
\end{myex}
\begin{mylm}
If $X(\tau)\in \mathcal{X}_0$ at time $\tau>0$, then $X(t)\in \mathcal{X}_0$, $\forall t\geq \tau$.
\end{mylm}
\begin{IEEEproof}
Assume that the state vector $X(\tau)$ is in $\mathcal{X}_0$. If two nodes $i$ and $j$ get in contact with each other at any time $t>\tau$, then according to Definition \ref{def1}, the outputs of transition, i.e. the sets $\{v_i(t) \cap v_j(t)\}$ and $\{v_i(t)\cup v_j(t)\}$ would be $\{v_i(t)\},\{v_j(t)\}$ or $\{v_j(t)\},\{v_i(t)\}$. Since we know that $X(\tau)\in \mathcal{X}_0$, $X(t)$ is also in $\mathcal{X}_0$. Thus, the proof is complete.
\end{IEEEproof}

\begin{mydef}
 The Lyapunov function $V(X(t))$ is defined as follows:
\begin{equation}
V(X(t))=nK^2-\displaystyle\sum_{i=1}^n |v_i(t)|^2.
\end{equation}
\end{mydef}

\begin{mylm}
If two nodes $i$ and $j$ contact with each other at time $t$, and $v_i(t)\not\subseteq v_j(t), v_j(t)\not\subseteq v_i(t)$, then there would be a reduction in the Lyapunov function $V(X(t))$.
\label{lm2}
\end{mylm} 
\begin{IEEEproof}
Suppose that $|v_i(t)|=l$, $|v_j(t)|=l^{\prime}$, and $|v_i(t)\cap v_j(t)|=r$. The change in the Laypunov function is:
\begin{equation}
V(X(t^{+}))-V(X(t))=-[(l+l^{\prime}-r)^2+r^2]+(l^2+l^{\prime^2})\leq -1,
\end{equation}
which is strictly less than zero for all $0\leq r\leq \min(l,l^{\prime})-1$. 
\end{IEEEproof}

\begin{mydef} Let $X(0)=x$. We denote the time that the state vector hits the set $\mathcal{X}_0$ for the first time by $\tau_x$, i.e.:
\begin{equation}
\tau_x=\min\{t>0|X(t)\in \mathcal{X}_0\}.
\end{equation}
\end{mydef} 

\begin{mydef}
Let $Y(t)$ be a random walk on the network starting from node $p$. We define $T_{pq}$ to be the first time $Y(t)$ visits node $q$:
\begin{equation}
T_{pq}=\min\{t\geq 0\mid Y(0)=p,Y(t)=q\},
\end{equation}
and the worst case hitting time, $\sigma$, is defined as follows:
\begin{equation}
\sigma=\max_{p,q\in V} \mathbb{E}\{T_{pq}\}.
\label{eq1}
\end{equation}
\end{mydef}

\begin{mylm} There exists $\epsilon>0$ such that:
\begin{equation}
\mathbb{E}\{V(X(t+2\sigma))-V(X(t))\mid X(t)=y\}\leq -\epsilon, \mbox{if } y\not \in \mathcal{X}_0,\\
\end{equation}
where $\sigma$ is defined in (\ref{eq1}).
\label{lm1}
\end{mylm}

\begin{IEEEproof}
Consider two random walks $Y(t)$ and $Z(t)$ on the network starting from nodes $p$ and $q$, respectively. We define the coalescing time of two random walks $Y(t),Z(t)$  as follows:
\begin{equation}
C_{pq}=\min\{t\geq 0 \mid Y(t)=Z(t), Y(0)=p, Z(0)=q\}.
\end{equation}

It follows from Markov inequality that:
\begin{equation}
\min_{p,q\in V} \Pr\{C_{pq}\leq t_0\}\geq 1-\frac{\sigma}{t_0}.
\label{eq11}
\end{equation}

If $X(t)\not\in \mathcal{X}_0$, then there exist $v_i(t)$, $v_j(t)$ such that $v_i(t)\not\subseteq v_j(t), v_j(t)\not\subseteq v_i(t)$. The corresponding random walks meet each other (or some other intermediate such value sets) up to time $t+2\sigma$ with probability at least $\frac{1}{2}$, which results in reduction of $V(X(t))$ according to Lemma \ref{lm2}. Otherwise, $X(t)$ is in the set $\mathcal{X}_0$. 

\end{IEEEproof}

\begin{mylm} The state vector $X(t)$ hits the set $\mathcal{X}_0$ in bounded time with probability one.
\label{lm4n}
\end{mylm}
\begin{IEEEproof}
We know that the Lyapunov function $V(X(t))$ is lower bounded by zero. According to Foster's criteria (see \cite{asmussen2003applied}, page 21) and Lemma \ref{lm1}, it follows that:
\begin{equation}
\Pr(\tau_x <\infty)=1, \forall x\not \in \mathcal{X}_0.
\end{equation}
\end{IEEEproof}

\subsection{Dissemination of Result in Memories}
In this part, we show how memory updating disseminates the correct result in the network. Without loss of generality, in the remainder of this paper, we assume that $\#c_1>\#c_2>\cdots>\#c_K$.\footnote{We assume that $\#c_K>0$. Otherwise, we can reduce the problem to the case with fewer choices. By slight modifications, the DMVR algorithm can also work for the cases where we have choices with equal size.}
\begin{mydef}
Assume that the state vector $X(t)$ gets in $\mathcal{X}_0$ at time $\tau$. We define the vector $v^{\star}=[v^1,\cdots,v^K]$ as follows:
\begin{equation}
v^k=
\begin{cases}
v_i(\tau),\qquad \exists i\in\{1,\cdots,n\} : |v_i(\tau)|=k,\\
\emptyset, \qquad\qquad\mbox{ otherwise.} 
\end{cases}
\end{equation}
and $r_k(t)=\big\vert\left\{ i \big\vert |v_i(t)|=k\right\}\big\vert$.
\end{mydef}

\begin{myth}
The vector $[\pi_1,\cdots,\pi_K]$ defined in (\ref{eq2}), gives the correct ranking of choices in a finite time with probability one.
\label{mainth} 
\end{myth}
\begin{IEEEproof}
 From Lemma \ref{lm4n}, we know that the state vector $X(t)$ eventually gets in the set $\mathcal{X}_0$ at time $\tau_x$. For a choice $c_k$, let $\alpha(k)$ be the smallest index such that $c_k\in v^{\alpha(k)}$. According to the definition of the convergence set and the preservation property, we know that: $\#c_k=\displaystyle\sum_{i=\alpha(k)}^K r_i(t)$. Hence, we have: $\#c_k>\#c_{k^{\prime}} \Longleftrightarrow \alpha(k)<\alpha(k^{\prime})$. Thus, based on assumption $\#c_1>\#c_2>\cdots>\#c_K$, the only possibility is: $r_k(t)>0$ and $\alpha(k)=k$, $\forall k\in\{1,\cdots,K\}$ for all $t>\tau_x$. Since $r_k(t)>0$, there exists at least one value set $v^k$, $1\leq k\leq K$ in the network. The value sets $v^k$ take random walk in the network and set the memories $m_{i,k}(t)$ to $v^k$ for all $i\in\{1,\cdots,n\}$. Let $\tau^{\prime}>\tau_x$ be the time that $m_{i,k}(t)=v^k$, $\forall i\in \{1,\cdots,n\}, 1\leq k\leq K$. Based on the definition of $\alpha(k)$, we have: $c_k=v^{\alpha(k)}\backslash v^{\alpha(k)-1}=m_{i,k}(t)\backslash m_{i,k-1}(t)$. Hence, all nodes obtain the correct ranking after time $\tau^{\prime}$.
\end{IEEEproof}
From above theorem, $m_{i,1}(t)$ gives the choice in majority. Hence, the DMVR algorithm can solve the majority voting problem by just updating $m_i(t)=m_{i,1}(t)$.

\begin{myrm} In the proposed solution, each node can also vote for more than one choice. To do so, it is sufficient to initialize the value set of each node to the union of its preferred choices. In this general case as well, the DMVR algorithm gives the correct ranking based on the size of choices. 
\end{myrm}

\subsection{State-optimal Implementation of DMVR Algorithm}
For the case of majority voting, the state of node $i$ is the pair $(m_i(t),v_i(t))$ where the sets $m_i(t)$ and $v_i(t)$ have $K$ and $2^K$ possible states, respectively. Thus, the total number of states is $K\times 2^K$. However, we can implement the DMVR algorithm with fewer states by adding the following rules:
\begin{itemize}
\item If the output of updating rule, $v_i(t)\cap v_j(t)$, is the empty set, we replace it by the set $\mathcal{C}$.
\item 
If $m_i(t^+)\not\subseteq v_i(t^+)$, we select a random member $c_k\in v_i(t^+)$ and change $m_i(t^+)$ to $\{c_k\}$. Otherwise, $m_i(t^+)$ remains unchanged. Then, the state of node $i$ is saved in the form of $\big(m_i(t^+), v_i(t^+)\backslash m_i(t^+)\big)$.
\end{itemize}

When $X(t)\in \mathcal{X}_0$, there is at least one value set $\{c_1\}$ in the network. When this value set meets a new node with $m_i(t)\neq \{c_1\}$, it updates $m_i(t)$ to $\{c_1\}$ and $m_i(t)$ will never be changed after that. Because $\{c_1\}\subseteq v_i(t), \forall i\{1,\cdots,n\}$ when $X(t)\in \mathcal{X}_0$.
For a fixed $m_i(t^+)$, the number of possible states for $v_i(t^+)\backslash m_i(t^+)$ is $2^{K-1}$. Consequently, the total number of states reduces to $K\times 2^{K-1}$. As an example, in the ternary voting, we have the following 12 states:
\begin{equation}
\begin{split}
(\{c_1\}, \emptyset), (\{c_1\},\{c_2\}), (\{c_1\},\{c_3\}), (\{c_1\},\{c_2,c_3\})\\
(\{c_2\}, \emptyset), (\{c_2\},\{c_1\}), (\{c_2\},\{c_3\}), (\{c_2\},\{c_1,c_3\})\\
(\{c_3\}, \emptyset), (\{c_3\},\{c_1\}), (\{c_3\},\{c_2\}), (\{c_3\},\{c_1,c_2\})
\end{split}
\end{equation} 

Thus, the number of states for ternary voting is 12 compared to 15 for the PAGA automaton \cite{Benezit2011} and it is equal to 32 for quaternary voting while the number of states for the PAGA is 100. 

In the case of ranking problem, we replace the value set and memories by an ordered $K$-tuple $a_i(t)=[a_i^1(t),\cdots,a_i^K(t)]$, which is a permutation of the set $\mathcal{C}$ along with an integer $1\leq p_i(t)\leq K$ which we perceive as a pointer to an entry of $a_i(t)$. At the beginning of the algorithm, each node $i$ puts its preferred choice in the first entry of $a_i(0)$, and sets an arbitrary permutation of other choices in the remaining entries. It also sets $p_i(0)=1$. Let $a_i^{1:p_i(t)}$ be a $p_i(t)$-tuple containing the first $p_i(t)$ entries of $a_i(t)$ and $\{a_i^{1:p_i(t)}\}$ be the set representation of it without any order.  
For a set $A\subseteq\mathcal{C}$, we define $\Pi_{A,a_i(t)}$ as a permutation of $A$ that preserves the order of entries in accordance with $a_i(t)$. Now, assume that two nodes $i$ and $j$ get in contact with each other at time $t$, and let $p_i(t)\leq p_j(t)$. Then, we apply the following updating rule:
\begin{equation}
\begin{split}
 a_i(t^+) &:=[\Pi_{A_1, a_i(t)}, \Pi_{A_2\backslash A_1, a_i(t)}, \Pi_{\mathcal{C}\backslash A_2, a_i(t)}],\\
a_j(t^+) &:=[\Pi_{A_1, a_j(t)}, \Pi_{A_2\backslash A_1, a_j(t)}, \Pi_{\mathcal{C}\backslash A_2, a_j(t)}],
\label{equ}
\end{split}
\end{equation}
where $A_1=\{a_i^{1:p_i(t)}\}\cap\{a_j^{1:p_j(t)}\}$, $A_2=\{a_i^{1:p_i(t)}\}\cup\{a_j^{1:p_j(t)}\}$. We also set $p_i(t^+):= |A_2|$ and $p_j(t^+):=|A_1|$. It is not difficult to verify that this form of implementing the DMVR algorithm can solve the ranking problem by the same arguments in Theorem \ref{mainth}.

From the above implementation, we can run the DMVR algorithm for the ranking problem with $K\times K!$ states where the terms $K$ and $K!$ are the possible values of the pointer $p_i(t)$ and the ordered tuple $a_i(t)$, respectively. 
\begin{myex}Suppose that $\mathcal{C}=\{c_1,c_2,c_3,c_4\}$. Two examples for the updating rule in (\ref{equ}) are given as follows:

\begin{align}
\nonumber  a_i(t)=[\overset{\downarrow}{1}&,3,2,4],  a_j(t)=[2,\overset{\downarrow}{1},4,3]
 \\ \nonumber&\longrightarrow a_i(t^+)=[1,\overset{\downarrow}{2},3,4], a_j(t^+)=[\overset{\downarrow}{1},2,4,3],\\
 \nonumber a_i(t)=[\overset{\downarrow}{1}&,4,2,3],  a_j(t)=[\overset{\downarrow}{3},1,2,4]
 \\\nonumber &\longrightarrow a_i(t^+)=[1,\overset{\downarrow}{3},4,2], a_j(t^+)=[3,1,2,\overset{\downarrow}{4}],
 \end{align}
where the arrow symbol points the $p_i(t)$-th entry of $a_i(t)$. 
\end{myex}
The following theorem proposes a lower bound for the required number of states of any ranking algorithm. This bound meets the required number of states of the DMVR algorithm, proving optimality of it for the ranking problem.
\begin{myth}
Any algorithm that finds the correct ranking in a finite time with probability one over arbitrary network topology, requires at least $K\times K!$ number of states per node. 
\label{TheoremApp}
\end{myth}
\begin{IEEEproof}
It is enough to show that the theorem applies in complete graphs. Suppose that the ranking problem can be solved by running algorithm $\mathcal{A}$ at each node in a finite time. Consider that each node $i$ has a state $s_i(t)$. Let integer $r$ corresponds to the ranking $[\pi_1,\cdots,\pi_K], 1\leq r\leq K!$. We define a class of states, $\mathcal{S}_r$, as a set of states which nodes associate with ranking $r$. Algorithm $\mathcal{A}$ is said to converge to ranking $r$ at time $\tau$ if $s_i(t)\in \mathcal{S}_r ,\forall i \in\{1,\cdots,n\}, \forall t\geq \tau$. We denote members of the class $\mathcal{S}_r$ by $\mathcal{S}_r=\{\mathcal{S}_r^1,\cdots,\mathcal{S}_r^{D_r}\}$ where $D_r=|\mathcal{S}_r|$. Let $N_1$ be the set of initial configurations that are consistent with the ranking $r=[\pi_1,\cdots, \pi_K]$, i.e.:
 \begin{align}
\nonumber& N_1=
 \\&\Big\{[\#\pi_1,\cdots,\#\pi_K]\in \mathbb{Z}_{+}^{K}:\displaystyle\sum_{i=1}^K \#\pi_i=n; \#\pi_i> \#\pi_j, \forall i>j\Big\}.
 \end{align}

 Let $n_i, 1\leq i\leq D_r$ be the number of nodes in $\{1,\cdots,n\}$ whose state is $\mathcal{S}_r^i$. In complete graphs, the whole information of $[s_1(t),\cdots,s_n(t)]$ can be represented by the set $[n_1,\cdots,n_{D_r}]$. Thus, the class $\mathcal{S}_r$ corresponds to a subset of the following set:
\begin{equation}
 N_2=\Big\{[n_1,\cdots,n_{D_r}]\in \mathbb{Z}_{+}^{D_r}:\displaystyle\sum_{i=1}^{D_r} n_i=n\Big\}.
 \end{equation}
 
Let $B_z$ be the set of vectors $[n_1,\cdots,n_{D_r}]$ that are achievable from the initial configuration $z\in N_1$.
\begin{mylm}
Consider two initial configurations $x,y\in N_1$, $x\neq y$. Then, we have: $B_x\cap B_y=\emptyset$.
\label{applm}
\end{mylm}
\begin{IEEEproof}
By contradiction. Suppose that there exists $x_0\in B_x\cap B_y$. We run algorithm $\mathcal{A}$ in a complete graph with nodes $\{1,\cdots,n+n^{\prime}\}$ for two different initial configurations $x,y$ of nodes $\{1,\cdots,n\}$. The algorithm $\mathcal{A}$ should give the correct result for any scheduling of local clocks at nodes. Consider a scheduling which only clocks of nodes in the set $\{1,\cdots,n\}$ are ticking up to time $T$ and the states of these nodes become $x_0$. Now, suppose that a centralized algorithm $\mathcal{A}_c$ want to obtain correct ranking by just looking at states of nodes in $\{1,\cdots,n\}$ and votes of nodes in $\{n+1,\cdots,n+n^{\prime}\}$. Algorithm $\mathcal{A}_c$ finds the ranking of votes of nodes in $\{1,\cdots,n\}$ by the vector $x_0$. But the centralized algorithm still needs to obtain all differences
$\#\pi_i-\#\pi_{i+1}$, $1\leq i\leq K-1$. Otherwise, it cannot rank votes of nodes in the whole network correctly. However, the two different initial configurations $x,y$ are mapped to the same state vector $x_0$. Consequently, the algorithm $\mathcal{A}_c$ cannot recover the correct initial configuration which is a contradiction. Thus, the proof of lemma is complete.
\end{IEEEproof}

We know that $|N_2|={n+D_r-1 \choose D_r-1}=\Theta(n^{D_r-1})$. Furthermore, we have:
\begin{equation}
\begin{split}
 &|N_1|=
 \\&=\frac{1}{K!}\Big|\Big\{[\#\pi_1,\cdots,\#\pi_K]\in \mathbb{Z}_{+}^{K}:\displaystyle\sum_{i=1}^K \#\pi_i=n, \#\pi_i\neq \#\pi_j\Big\}\Big|\\
&\geq \frac{1}{K!}\Big(\Big|\Big\{[\#\pi_1,\cdots,\#\pi_K]\in \mathbb{Z}_{+}^{K}:\displaystyle\sum_{i=1}^K \#\pi_i=n\Big\}\Big|-\\
&\displaystyle\sum_{j,k} \Big|\Big\{[\#\pi_1,\cdots,\#\pi_K]\in \mathbb{Z}_{+}^{K}:\displaystyle\sum_{i=1}^K \#\pi_i=n, \#\pi_j= \#\pi_k\Big\}\Big|\Big)\\
&={n+K-1 \choose K-1}-{K \choose 2}\times {n+K-2 \choose K-2}=\Theta(n^{K-1}).
\end{split}
\end{equation}


Let $F:N_1\longrightarrow N_2$ which maps $x\in N_1$ to $B_x\subset N_2$. According to Lemma \ref{applm}, the mapping $F$ should be invertible. For sufficiency large $n$, this can occur only if $D_r\geq K$. Consequently, it can be concluded that each class $\mathcal{S}_r$ has at least $K$ members and total number of states is at least $K\times K!$.
\end{IEEEproof}
\section{Time Complexity}
In this section, we first analyze the time complexity of the DMVR algorithm for the binary voting problem in complete graphs. Then, we study the multiple choice case and derive a tight bound on the running time of the DMVR algorithm for the ranking problem. At the end, we propose a method to speed up the DMVR algorithm in majority voting problem.
\subsection{Binary Voting Case}
In order to study the time complexity of the DMVR algorithm, we divide its execution time into two phases:
\begin{itemize}
\item First phase (extinction of $\{c_2\}$): This phase starts at the beginning of the algorithm and continues until none of the value sets are $\{c_2\}$. We denote the finishing time of this phase by $\tau_1$. 
\item Second phase (dissemination of $\{c_1\}$ in memories): This phase follows the first phase and it ends when memories of all nodes are $\{c_1\}$. The execution time of this phase is represented by $\tau_2$.
\end{itemize} 
\subsubsection{Time complexity of the first phase}
For the binary case, the transition rule of DMVR algorithm is exactly the same as the PAGA algorithm. In \cite{draief2012convergence}, an upper bound, $O(\log(n)/(1-2\rho))$, is given for the PAGA algorithm in complete graphs where $\rho_2=\rho$. Here, we propose exact average time complexity for the first phase. Suppose that the number of nodes voting for $c_1$, $c_2$ are $s=n(1-\rho)$ and $r=n\rho$ at the beginning of the algorithm. We denote the sets of nodes having value sets $\{c_1\}$ and $\{c_2\}$ at time $t$ by $S_1(t)$ and $S_2(t)$, respectively. Consider the Markov chain in Fig. \ref{figma1}. The state $r-i$, $0\leq i\leq r$, represents the number of nodes whose value sets are $\{c_2\}$. Suppose that the state of Markov chain is $r-i$ at time $t$. The chain undergoes transition from state $r-i$ to $r-i-1$ if one of nodes in the set $S_2(t)$ gets in contact with one of nodes in $S_1(t)$, which occurs with rate $2(r-i)\times(s-i)/n$.  After updating the value sets, both $|S_1(t)|$ and $|S_2(t)|$ will decreased exactly by one. Let $T_{r-i}^1$ be the sojourn time in state $r-i$. Hence, we have:

\begin{figure}[!t]
\centering
\includegraphics[width=3in]{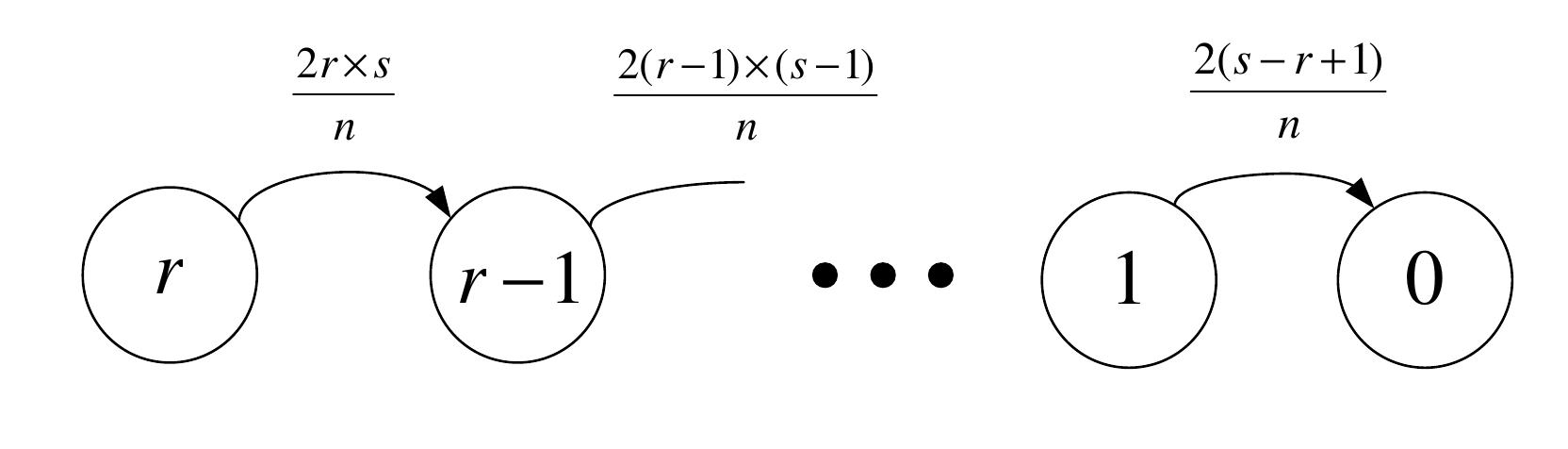}
\caption{The Markov chain model for the first phase. State $r-i$ shows the size of the set $S_2(t)$.}
\label{figma1}
\end{figure}

\begin{equation}
\begin{split}
\mathbb{E}\{\tau_1\}=\displaystyle\sum_{i=0}^{r-1} \mathbb{E}\{T^1_{r-i}\}&=\displaystyle\sum_{i=0}^{r-1} \frac{n}{2}\times \frac{1}{(r-i)(s-i)}
\\&\approx\frac{n}{2(s-r)} \log(\frac{r(s-r)}{s}),
\end{split}
\label{eqe}
\end{equation}
in terms of time units. Thus, the average running time of the first phase is: $\mathbb{E}\{\tau_1\}\approx\frac{1}{2(1-2\rho)} \log(\frac{n\rho(1-2\rho)}{1-\rho})$ time units. Furthermore, we can obtain the variance of $\tau_1$ as follows:
\begin{align}
\mathrm{Var}(\tau_1)&=\displaystyle\sum_{i=0}^{r-1} \mathrm{Var}\{T^1_{r-i}\}=\displaystyle\sum_{i=0}^{r-1} \frac{n^2}{4}\times \frac{1}{(r-i)^2(s-i)^2}.
\label{eqv}
\end{align}

\subsubsection{Time complexity of the second phase}
At the beginning of the second phase, the number of nodes in $S_1(t)$ is $n(1-2\rho)$ and all the remaining nodes have the value sets $\{c_1,c_2\}$ or $\emptyset$. Furthermore, the memories of all of these nodes are $\{c_2\}$ in extreme case. Consider the Markov chain in Fig. \ref{figma2}. The state $r-i, 0\leq i\leq r$, represents the number of nodes having value sets $\{c_1,c_2\}$ or $\emptyset$ with memory $\{c_2\}$. We denote the set of such nodes by $\mathcal{M}_2(t)$. There is a reduction in $|\mathcal{M}_2(t)|$ if and only if a node in $\mathcal{M}_2(t)$ gets in contact with a node in $S_1(t)$. If $|\mathcal{M}_2(t)|=r-i$, then with rate $2\times \frac{(r-i)(n-2n\rho)}{n}$, there is a transition from state $r-i$ to state $r-i-1$. Let $T_{r-i}^2$ be the sojourn time in state $r-i$. Then, we have:
\begin{equation}
\begin{split}
\mathbb{E}\{\tau_2\}\leq\displaystyle\sum_{i=0}^{r-1} \mathbb{E}\{T^2_{r-i}\}&=\displaystyle\sum_{i=0}^{r-1} \frac{n}{2}\times \frac{1}{(r-i)(n-2n\rho)}
\\&\approx\frac{1}{2(1-2\rho)} \log(2n\rho),
\end{split}
\label{eqe2}
\end{equation}

Thus, we can conclude that the time complexity of the DMVR algorithm is:
\begin{equation}
\mathbb{E}\{\tau_1+\tau_2\}\leq\frac{1}{2(1-2\rho)}\times\Big(\log(\frac{n\rho(1-2\rho)}{1-\rho})+\log(2n\rho)\Big).
\end{equation} 
\subsection{Multiple Choice Voting Case}
Consider two choices $c_k$ and $c_l$. From the state vector $X(t)=[v_1(t),v_2(t),\cdots,v_n(t)]$, we define a new state vector $X^{k,l}(t)=[v^{\prime}_1(t),\cdots,v_n^{\prime}(t)]$ by projecting the value set of each node $i$ on $\{c_k,c_l\}$, i.e. $v_i^{\prime}(t)=v_i(t)\cap \{c_k,c_l\}$. Thus, the projected state vector $X^{k,l}(t)$ represents the path of execution in a binary voting with just two choices $c_k,c_l$. We define $\mathcal{X}_0^{k,l}$ to be the convergence set of the projected system as follows:
\begin{equation}
\begin{split}
X^{k,l}(t)\in \mathcal{X}^{k,l}_0 \mbox{ if } |v_i^{\prime}(t)|\leq |v_j^{\prime}(t)| \Longrightarrow v_i^{\prime}(t)&\subseteq v_j^{\prime}(t),\\
 &\forall i,j\in\{1,\cdots,n\}.
\end{split}
\end{equation}
\begin{mylm}
Let $\tau_x$ and $\tau_x^{k,l}$ be the time that the state vectors $X(t)$ and $X^{k,l}(t)$ hit their corresponding convergence sets. Then, we have: $\tau_x=\max_{k,l\in V, k\neq l} \tau_{x}^{k,l}$.
\end{mylm}
\begin{IEEEproof}
First, we prove that $\tau_x\geq\max_{k,l\in V, k\neq l} \tau_{x}^{k,l}$. If $X(t)\in \mathcal{X}_0$, then for all $k,l\in\{1,\cdots,K\}, k\neq l$:
\begin{equation}
\begin{split}
 \forall i,j\in & \{1,\cdots,n\}, |v_i(t)|\leq |v_j(t)| \Longleftrightarrow  v_i(t)\subseteq v_j(t) 
\\\Longrightarrow & |v^{\prime}_i(t)|\leq |v^{\prime}_j(t)| \mbox{ and } v^{\prime}_i(t)\subseteq v^{\prime}_j(t) \Longrightarrow X^{k,l}(t) \in \mathcal{X}^{k,l}_0.
\end{split}
\end{equation}

\begin{figure}[!t]
\centering
\includegraphics[width=3in]{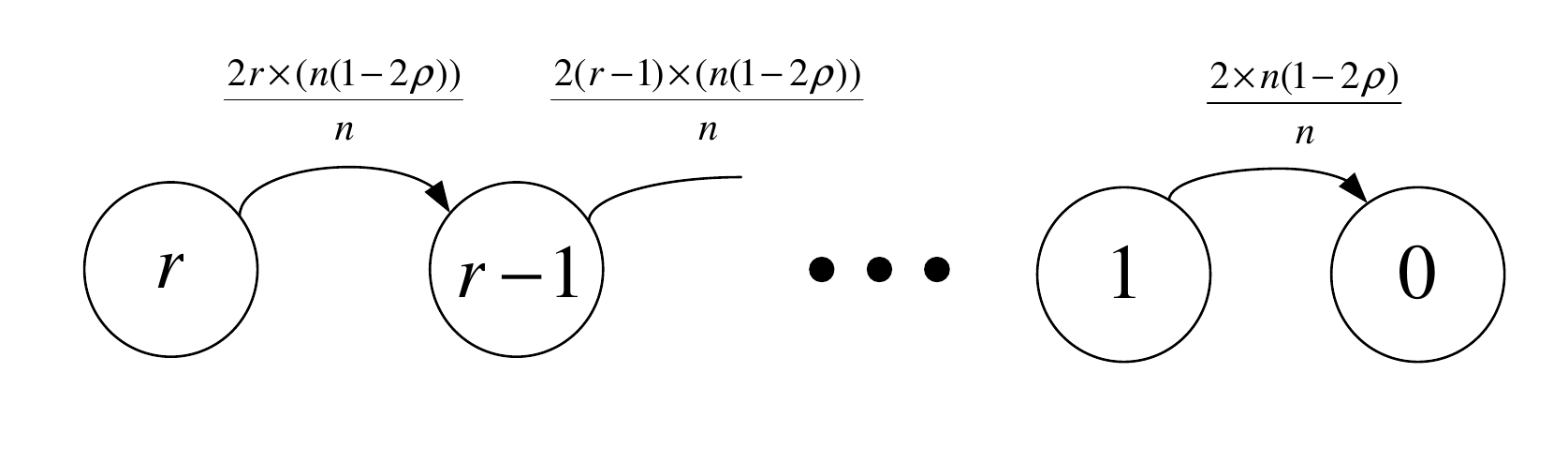}
\caption{The Markov chain model for the second phase. State $r-i$ shows the size of the set $\mathcal{M}_2(t)$.}
\label{figma2}
\end{figure}

Now, we show that $\tau_x\leq\max_{k,l\in V, k\neq l} \tau_{x}^{k,l}$. Consider any two nodes $i$ and $j$ at time $t=\max_{k,l\in V, k\neq l} \tau_{x}^{k,l}$. Without loss of generality, assume that $|v_i(t)|\leq |v_j(t)|$. We will show that $v_i(t)\subseteq v_j(t)$. By contradiction, suppose that there exists a choice $c_k$ such that $c_k\in v_i(t)$ and $c_k\not \in v_j(t)$. Now, consider any choice $c_l \in v_j(t)$. Since $t\geq \tau_{x}^{k,l}$, the state vector $X^{k,l}(t)$ has already hit its convergence set. This occurs if and only if $v_i(t)\cap \{c_k,c_l\}=\{c_k,c_l\}$ and $v_j(t)\cap\{c_k,c_l\}=\{c_l\}$. Hence, we can conclude that $c_l\in v_i(t), \forall c_l \in v_j(t)$. However, it means that $|v_i(t)|>|v_j(t)|$ which is a contradiction. 
\end{IEEEproof}
From (\ref{eqe}) and (\ref{eqv}), $\mathbb{E}\{\tau_{x}^{k,l}\}$ and $\mathrm{Var}(\tau_{x}^{k,l})$ can be obtained by substituting $r,s$ with $n\rho_k$ and $n\rho_l$, respectively.
\begin{mylm} (Order Statistics  \cite{bertsimas2006tight}) Let $[Z_1,\cdots,Z_R]$ denote $R\geq 2$ random variables (not necessarily independent or identically distributed) with means $[\mu_r]$ and variances $[\sigma^2_r]$. Let $Z_{max}=\max_{r=1,\cdots,R} Z_r$. Then, we have:
\begin{equation}
\mathbb{E}\{Z_{max}\}\leq\frac{1}{R}\displaystyle\sum_{r=1}^R \mu_r + \sqrt{\frac{R-1}{R}\displaystyle\sum_{r=1}^R \Big[\sigma_r^2 + (\mu_r-\frac{1}{R}\displaystyle\sum_{r=1}^R \mu_r)^2\Big] }
\end{equation}
\end{mylm}

 Now, we can give an upper bound on $\mathbb{E}\{\tau_x\}$ from above lemma:
\begin{equation}
\begin{split}
\mathbb{E}\{\tau_x\}&\leq\mu + \sqrt{\displaystyle\sum_{k,l\in V, k\neq l} \Big[\mathrm{Var} (\tau_x^{k,l}) + (\mathbb{E}\{\tau_x^{k,l}\}-\mu)^2\Big] }
\\&=O(\frac{\log(n)}{\underset{j=1,\cdots,K-1}{\min}\rho_{j+1}-\rho_j}),
\end{split}
\label{eqb1}
\end{equation}
where $\mu=\frac{1}{{K \choose 2}} \displaystyle\sum_{k,l\in V, k\neq l} \mathbb{E}\{\tau_x^{k,l}\} $.

After the state vector gets in the convergence set, we should still wait to copy vector $v^{\star}$ in memories of all nodes. At time $\tau_x$, the number of nodes with the value state $\{c_1,c_2,\cdots,c_j\}$ is $n\rho_j-n\rho_{j+1}$, $1\leq j<K$. Let $\mathcal{M}_j^{\prime}$ be the set of such nodes and $\tau_j^{\prime}$ be the time until memories $m_{i,j}(t)$'s of all nodes are set to $\{c_1,\cdots,c_j\}$. When a node $i$ contacts with any node in $\mathcal{M}_j^{\prime}$, its memory $m_{i,j}(t)$ will be set to $\{c_1,\cdots,c_j\}$. With the same arguments in previous part, we have:
\begin{equation}
\begin{split}
\mathbb{E}\{\tau_j^{\prime}\}\leq \frac{n}{2} \displaystyle\sum_{i=0}^{n-1} \frac{1}{(n-i)(n\rho_j-n\rho_{j+1})}\approx \frac{1}{2(\rho_j-\rho_{j+1})} \log(n),
\\ \mathrm{Var} (\tau_j^{\prime})\leq \frac{n^2}{4} \displaystyle\sum_{i=0}^{n-1} \frac{1}{(n-i)^2(n\rho_j-n\rho_{j+1})^2}\approx\frac{1}{4(\rho_j-\rho_{j+1})^2}.
\end{split}
\end{equation}

Thus, an upper bound on $\tau^{\prime} =\max_{1\leq j<K} \tau_j^{\prime}$ can be obtained from order statistics:
\begin{equation}
\begin{split}
\mathbb{E}\{\tau^{\prime}\}&\leq \mu^{\prime}+\sqrt{\displaystyle\sum_{j=1}^{K-1} \Big[\mathrm{Var} (\tau_j^{\prime}) + (\mathbb{E}\{\tau_j^{\prime}\}-\mu^{\prime})^2\Big] }
\\&=O(\frac{\log(n)}{\underset{j=1,\cdots,K-1}{\min}\rho_{j+1}-\rho_j}),
\label{eqb2}
\end{split}
\end{equation}
where $\mu^{\prime}= \frac{1}{K-1}\displaystyle\sum_{j=1}^{K-1} \mathbb{E}\{\tau_j^{\prime}\}$. From the bounds in (\ref{eqb1}) and (\ref{eqb2}), we can conclude that the time complexity of the DMVR algorithm is $O(\frac{\log(n)}{\underset{j=1,\cdots,K-1}{\min}\rho_{j+1}-\rho_j})$.

\subsection{Speeding up the DMVR algorithm for majority voting problem}
The execution time of the DMVR algorithm can be divided into two phases: The first phase starts at time zero and it ends when the state vector $X(t)$ gets in the convergence set $\mathcal{X}_0$. Afterwards, the second phase starts and it terminates when all nodes' memories are set with the majority vote. In order to speed up the second phase, we add the following rule when two nodes $i$ and $j$ contact with each other at time $t$:

\begin{algorithmic}[1]
\If {$|v_i(t^+)|>1 \land |v_j(t^+)|> 1$}
\State Generate $u$ from Bernoulli distribution with success probability $0.5$.
\If {$u=1$}
\State $m_i(t^+):=m_j(t)$,
\Else
\State $m_j(t^+):=m_i(t)$.
\EndIf
\EndIf
\end{algorithmic} 

It is worth mentioning that the added rule is executed from the beginning of the algorithm. The idea behind this rule is that even nodes with the value sets other than the majority vote, cooperate in spreading the majority vote in the memories of all nodes. We call the proposed solution as the enhanced version of the DMVR algorithm. Simulation results show that the enhanced version of the DMVR algorithm can speed up the DMVR algorithm in complete graph, torus, and ring networks.

\begin{mylm}
Each node converges to the majority vote in a finite time with probability one by running the enhanced version of the DMVR algorithm.
\end{mylm}
\begin{IEEEproof}
Let $\tau_x$ be the time that the state vector $X(t)$ gets in the convergence set. Since then, the only value set with size one in the network would be $\{c_1\}$. Consider the vector $M(t)=[m_1(t),\cdots,m_n(t)]$. We define the Lyapunov function $V^{\prime}(M(t))$, $t>\tau_x$, as follows:
\begin{equation}
V^{\prime}(M(t))=n- |\{i\mid c_1\in m_i(t)\}|, t>\tau_x.
\end{equation}  

Suppose that two nodes $i$ and $j$ get in contact with each other at time $t>\tau_x$. Let $M_0$ be the vector of length $n$ with all entries equal to $\{c_1\}$. Then, we have:

\begin{align}
\nonumber &\mathbb{E}\{V^{\prime}(M(t^+))-V^{\prime}(M(t))\mid M(t)=y\}
  =\\&=\begin{cases}
\leq -\epsilon,\qquad &\mbox{ if } |v_i(t^+)|=1  \lor |v_j(t^+)|=1,\\
0\qquad\qquad  &\mbox{ if } |v_i(t^+)|>1  \land |v_j(t^+)|>1,
\end{cases}
\end{align}
where $y\neq M_0$. Hence, by the same arguments in Lemma \ref{lm1}, we conclude that $M(t)$ converges to vector $M_0$ with probability one.
\end{IEEEproof}
\section{Simulations}
\begin{figure}[!t]
\centering
\includegraphics[width=3.7in]{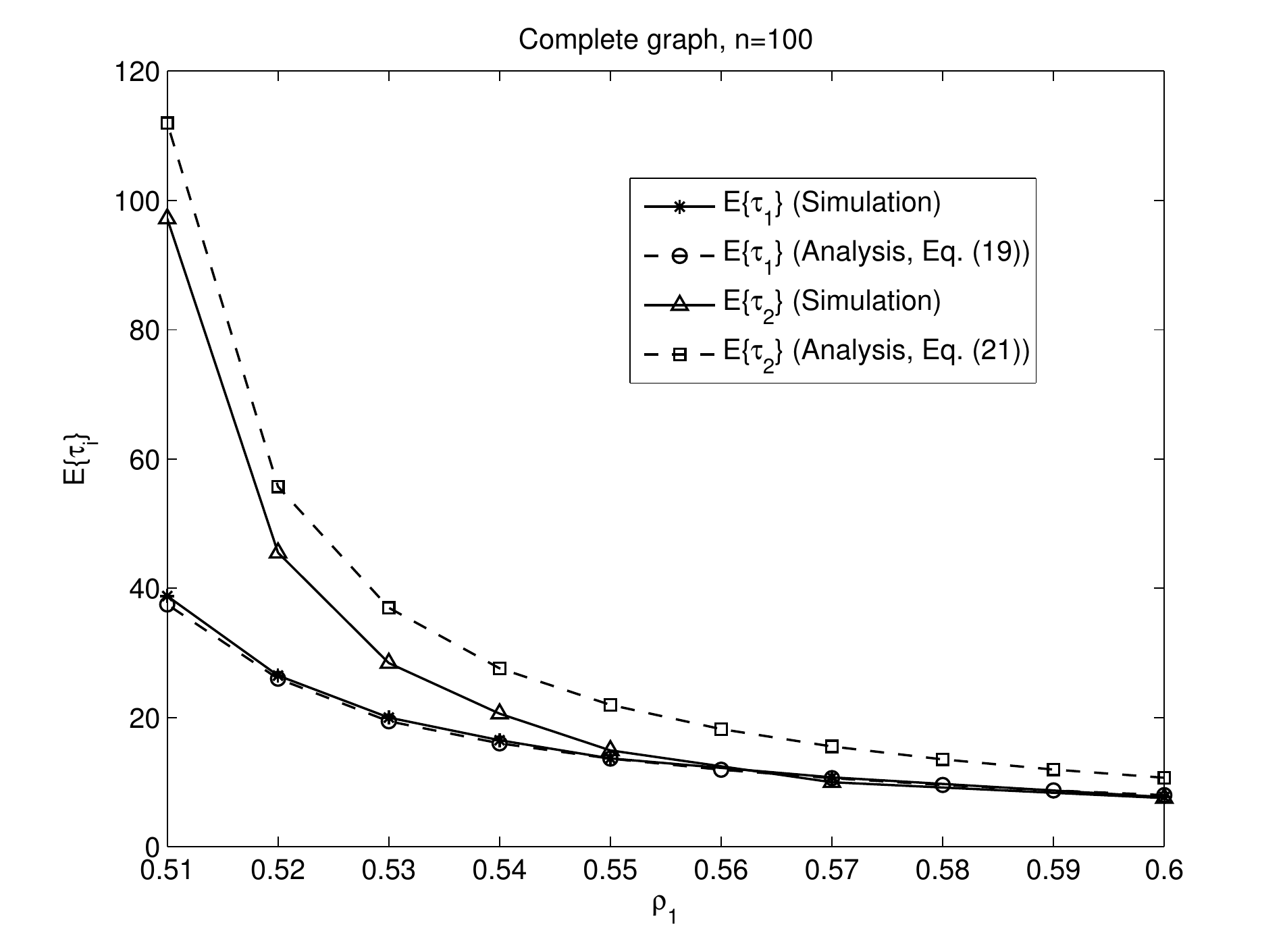}
\caption{Time complexities of first and second phases of the DMVR algorithm versus $\rho_1$ in binary voting, $n=100$.}
\label{figs1}
\end{figure}
In this section, we evaluate the time complexity of the DMVR algorithm through simulations and compare it with PAGA automaton in binary and ternary voting. Furthermore, we study  the proposed time complexity bounds in complete graphs. Each point in simulations is averaged over 1000 runs.

\begin{figure}[!t]
\centering
\includegraphics[width=3.7in]{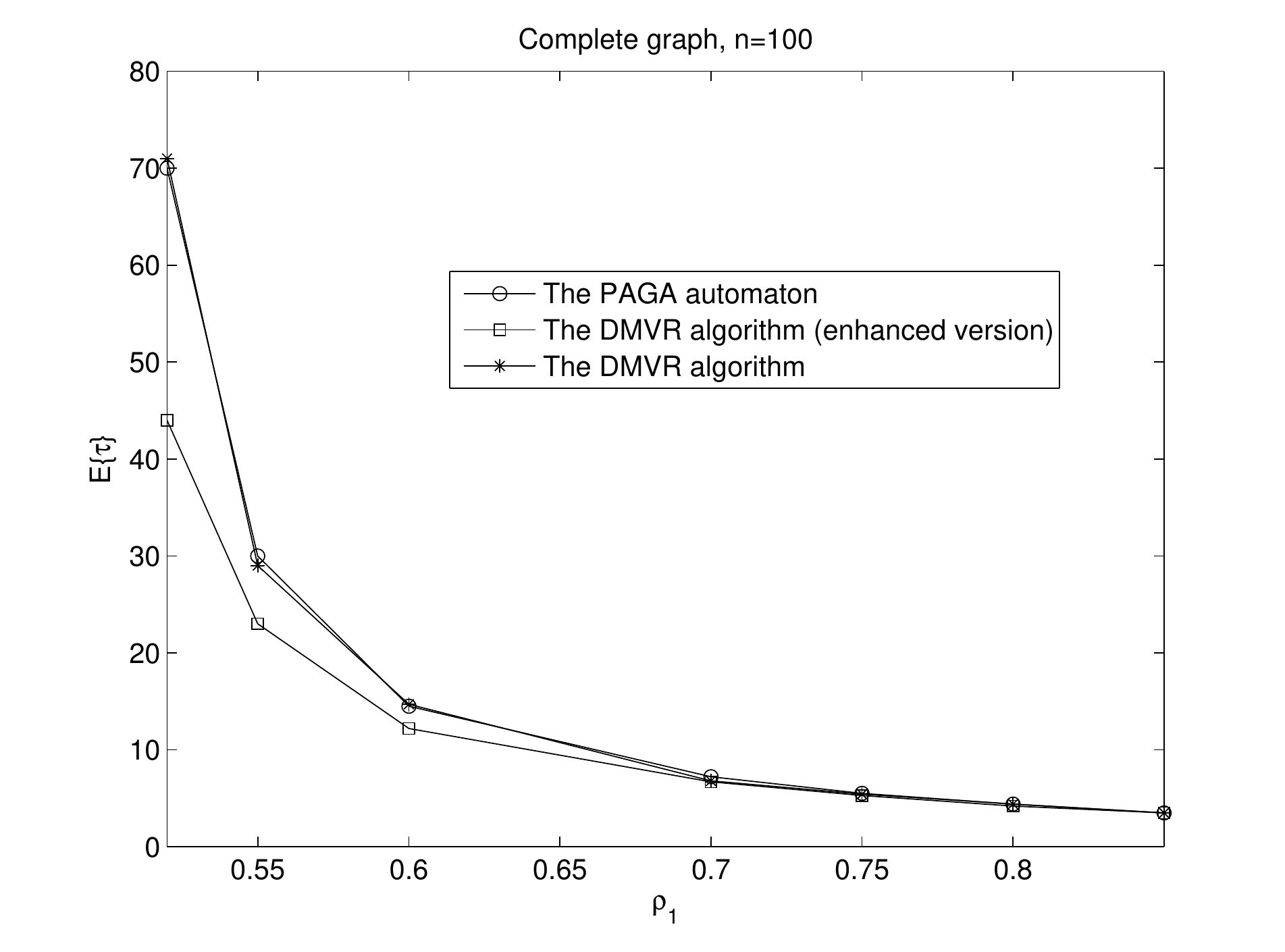}
\caption{Comparison of time complexities of the DMVR algorithm, its enhanced version, and the PAGA automaton for binary voting in complete graphs, $n=100$.}
\label{figs2}
\end{figure}

We compare the proposed bounds on $\mathbb{E}\{\tau_1\}$ and $\mathbb{E}\{\tau_2\}$ derived in (\ref{eqe}) and (\ref{eqe2}) with simulation results for binary voting in Fig. \ref{figs1}. As it can be seen, the bound $\mathbb{E}\{\tau_1\}$ is exact as we expected while there is a constant gap between simulation and analysis for $\mathbb{E}\{\tau_2\}$. 

\begin{figure}[!t]
    \centering
    \subfigure[Ring networks.]
    {
        \includegraphics[width=3.7in]{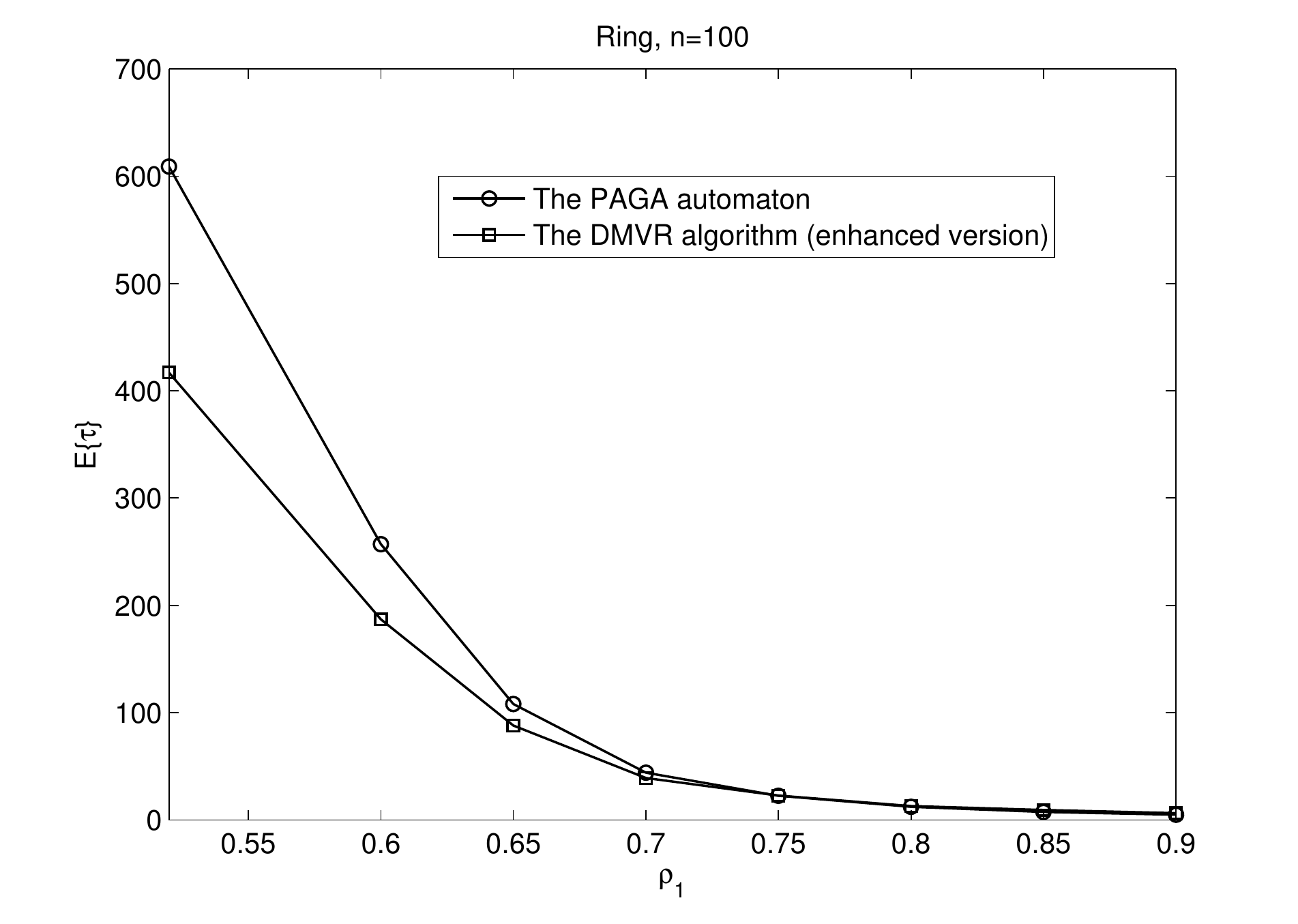}
        \label{figcompTa}
    }
    \subfigure[Torus networks.]
    {
        \includegraphics[width=3.7in]{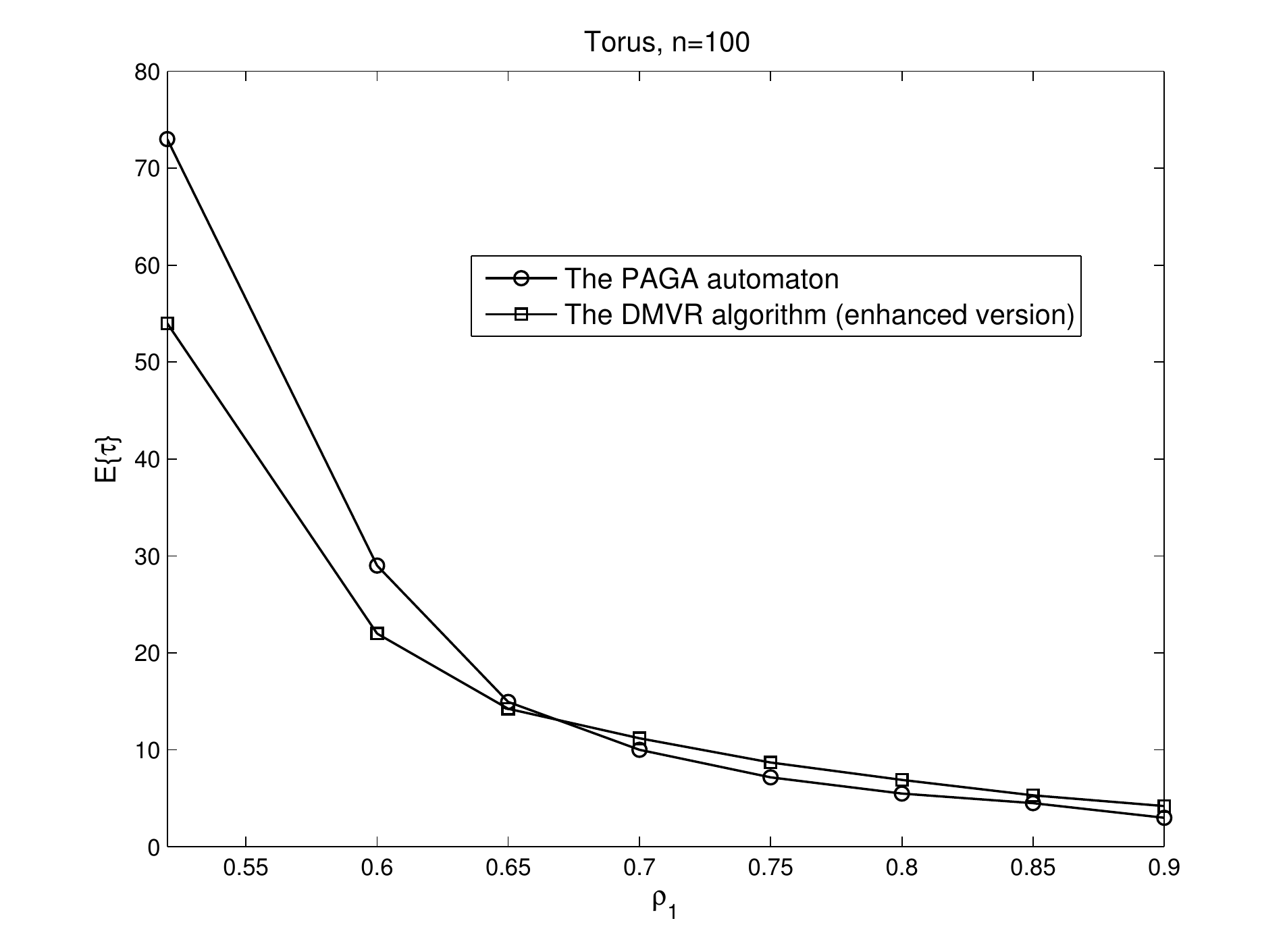}
        \label{figcompTb}
    }
	 \caption{Time complexities of enhanced version of the DMVR algorithm, and the PAGA automaton for binary voting in ring and torus networks, $n=100$.}
\label{figs3}
\end{figure} 

In Fig. \ref{figs2}, the time  complexities of DMVR algorithm, its enhanced version, and the PAGA automaton are depicted versus $\rho_1$. Since the transition rule of DMVR algorithm is identical to the PAGA automaton in binary case, the performance of two algorithms are very close to each other. However, the enhanced version of DMVR algorithm outperforms the other two algorithms as $\rho_1$ gets close to 0.5. In Fig. \ref{figs3}, we can also see this trend in ring and torus networks. 

\begin{figure}[!t]
\centering
\includegraphics[width=3.7in]{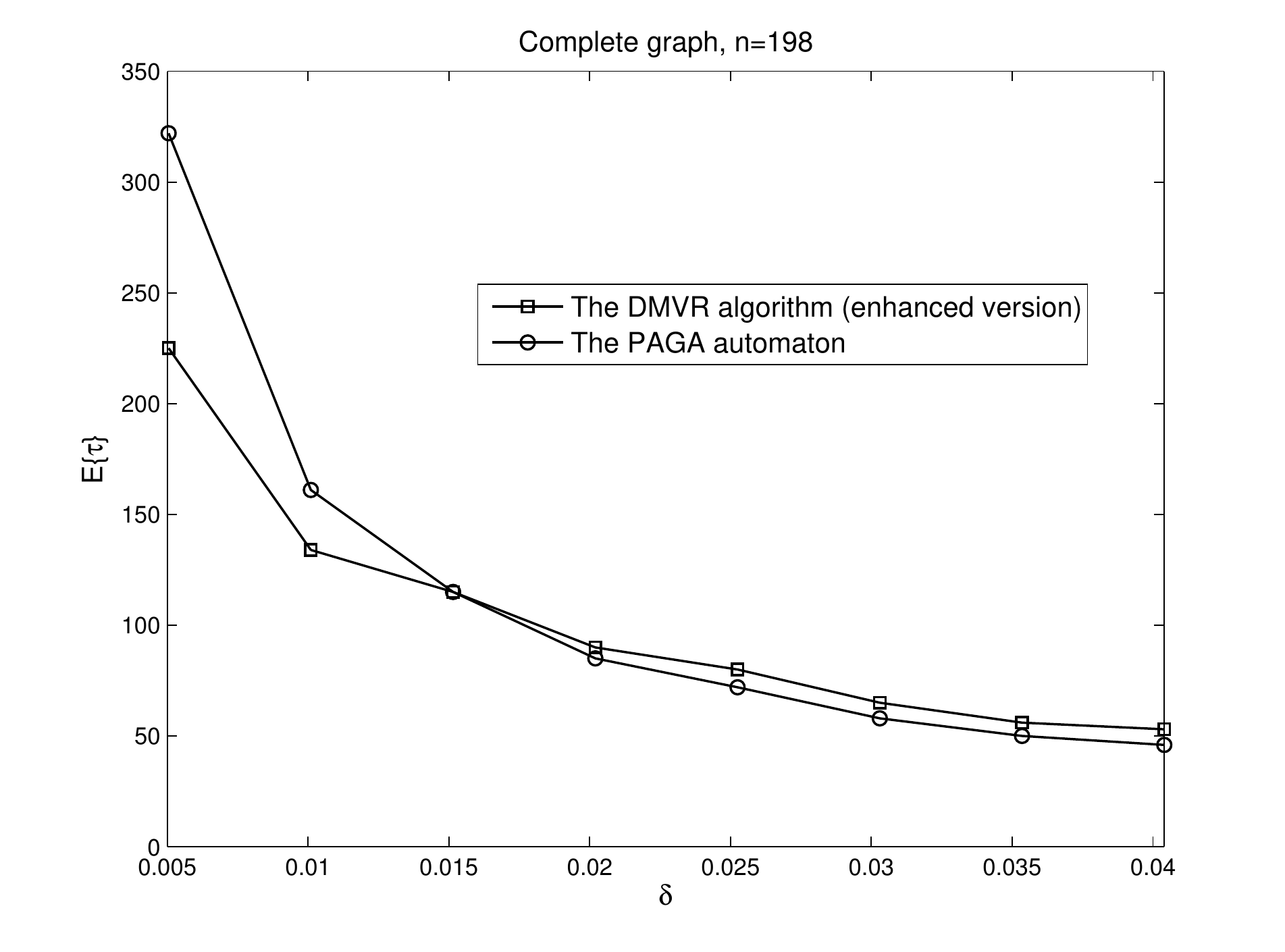}
\caption{Time complexities of enhanced version of the DMVR algorithm, and the PAGA automaton for ternary voting in complete graphs, $n=198$.}
\label{figs4}
\end{figure}

For ternary voting problem, we consider the percentage of initial votes in the form of $[\rho_1,\rho_2,\rho_3]=[\frac{1}{3}+\delta, \frac{1}{3}, \frac{1}{3}-\delta]$ where $0<\delta<\frac{1}{3}$. In Fig. \ref{figs4}, time complexities of enhanced version of the DMVR algorithm and the PAGA automaton are given for $\delta\in [0.005,0.041]$, $n=198$. As it can be seen, the enhanced version of the DMVR algorithm outperforms the PAGA automaton for small $\delta$.

\begin{figure}[!t]
\centering
\includegraphics[width=3.7in]{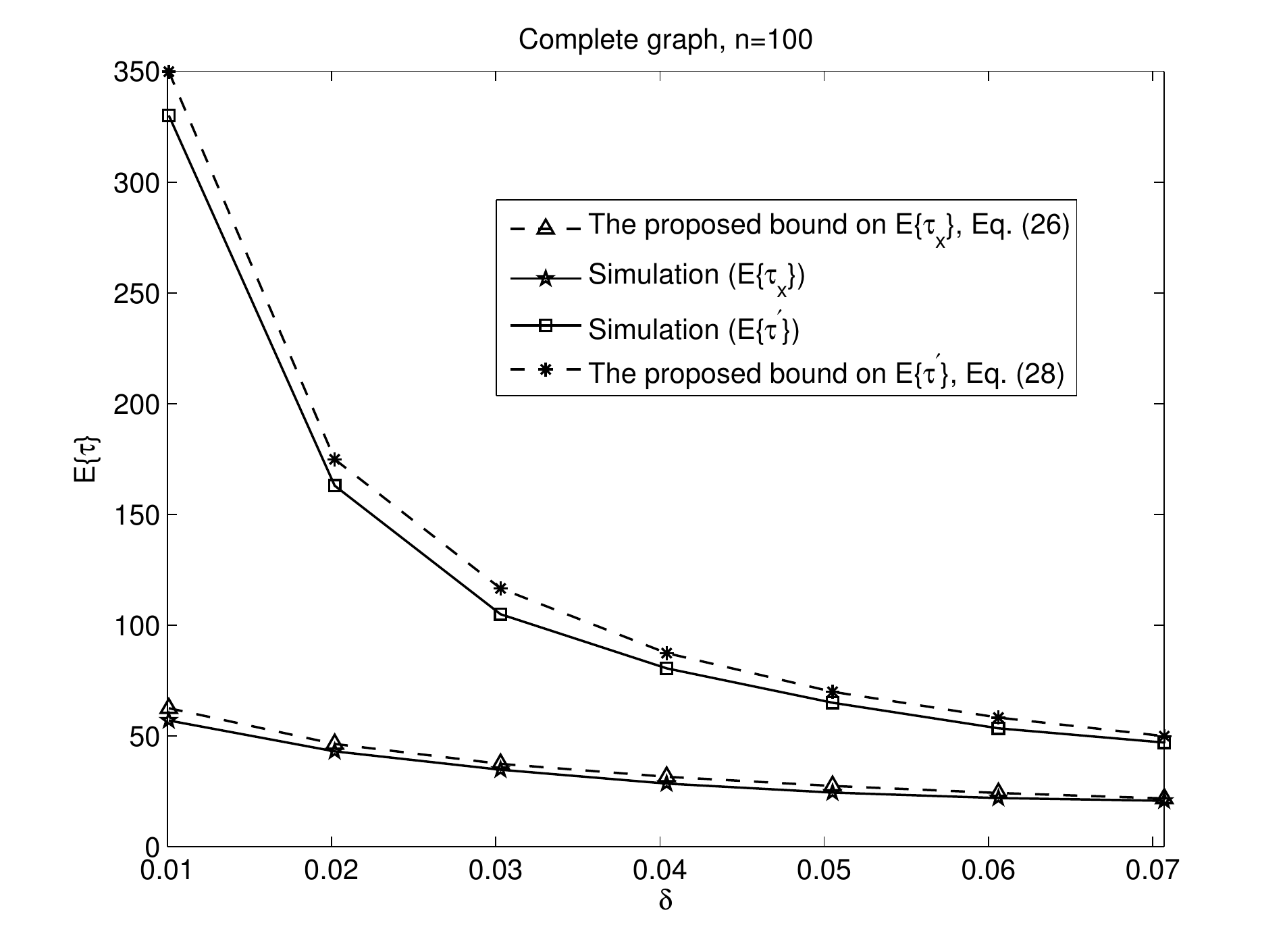}
\caption{Time complexity of the DMVR algorithm for the ranking problem with three votes in complete graphs, $n=100$.}
\label{figs5}
\end{figure}

 At the end, we compare the bounds for $\mathbb{E}\{\tau_x\}$ and $\mathbb{E}\{\tau^{\prime}\}$ derived in (\ref{eqb1}) and (\ref{eqb2}) for the ranking problem with three votes. In Fig. \ref{figs5}, bounds from order statistics have a small gap with simulation results and they can predict the behaviour of the DMVR algorithm accurately based on the percentage of initial votes.

\section{Conclusions}
In this paper, we proposed the DMVR algorithm in order to solve the majority voting and ranking problems for any number of choices. The DMVR algorithm is a simple solution with bounded memory and it is optimal for the ranking problem in terms of number of states. Furthermore, we analyzed time complexity of the DMVR algorithm and showed that it relates inversely to $\underset{i=1,\cdots,K-1}{\min} \rho_{i+1}-\rho_i$. As a future work, it is quite important to obtain the minimum required number of states for solving majority voting problem. We conjecture that the DMVR algorithm is an optimal solution for majority voting problem, i.e. at least $K\times 2^{K-1}$ states are required for any possible solution.

\bibliographystyle{IEEEtran}
\bibliography{IEEEabrv,mybib}

\end{document}